# Dynamic Safety Message Power Control in VANET Using PSO

Ghassan Samara
Department of Computer Science, Zarqa University
Zarqa, Jordan

Tareq Alhmiedat
Department of IT, Tabuk University
Tabuk , KSA.

Amer O. Abu Salem
Department of Computer Science, Zarqa University
Zarqa, Jordan

*Abstract*— In the recent years Vehicular Ad hoc Networks (VANET) became one of the most challenging research area in the field of Mobile Ad hoc Networks (MANET). Vehicles in VANET send emergency and safety periodic messages through one control channel having a limited bandwidth, which causes a growing collision to the channel especially in dense traffic situations. In this paper a protocol Particle swarm optimization Beacon Power Control (PBPC) is proposed, which makes dynamic transmission power control to adjust the transmission power of the safety periodic messages that have been aggressively sent by all vehicles on the road 10 times per a second, the proposed protocol aims to decrease the packet collision resulted from periodic safety messages, which leads to control the load on the channel while ensuring a high probability of message reception within the safety distance of the sender vehicle.

## I. INTRODUCTION

Creating an efficient safety system on the road is a very important and critical concern for humans today. Nearly 1.3 million people die as a result of road traffic accidents annually, and more than 3000 deaths each day are reported. More than half of the people involved in the accidents were not travelling in a vehicle [1]; moreover, the number of persons injured was 50 times greater than the number of recorded deaths each day [2]. The number of vehicles in 2004 is approximately 750 million globally [3], increasing annually by 50 million [4]. Today, the estimated number of vehicles exceeds one billion, increasing the possibility of more crashes and deaths on the roads. According to the World Health Organization WHO [2], road traffic accident is the fifth leading cause of death in the world, and each year, 2.4 million die from traffic related accidents [2]. Traffic congestion wastes time and fuel, thus, there is an urgent demand to develop efficient safety systems. The new techniques in this system should aim to make the intelligent vehicle think, communicate with other vehicles, and act to prevent accidents. To implement such a system, vehicle manufacturers have begun to equip their vehicles with devices enhancing safety, such as small range radars, night vision, light sensors, rain sensors, navigation systems, and the Event Data Record (EDR) resembling the Black-Box [5, 6].

Vehicles gain more fresh information when they communicate (talk) with each other and inform each other of any probable danger; they may even respond to that danger in a cooperative manner. However, VANET is still at the early stages of deployment, and real and intensive research pertaining to necessary safety solutions is still limited. This research gap prevents VANET from achieving its main goal of creating an efficient safety system on the road.

One of the major efforts dedicated to VANET was launched in 2011 where the United Nations (UN) Road Safety Collaboration has developed a global plan for the Decade of Action for Road Safety 2011–2020. The categories of activities include building road safety, improving the safety of road infrastructure, and broader transport networks; the plan also aims to develop safer vehicles and enhance the behavior of road users [2].

Wirelesses access in vehicular environment (WAVE) is a multi-channel approach, designed by the Federal Communications Commission (FCC), reserved for one control channel from 5.855 to 5865 GHz, for high availability, low latency vehicle safety communications [7]. Furthermore, WAVE represents the first VANET standard published in 2006. An enhancement was required on IEEE 802.11 standard to support applications from the Intelligent Transportation Systems (ITS), a branch of the U.S. Department of Transportation. The result showed the 802.11p standard, which was approved on July 2010 [8]. The 802.11p standard is meant for VANET communication and uses dedicated short range communications (DSRC) spectrum; it is divided into eight 10 MHz channels with only one control channel for safety application communication. VANET safety applications depend on the exchange of safety information among vehicles (C2C communication) or between vehicle to infrastructure (C2I Communication) using the control channel (see Figure 1). VANET safety communication is implemented in two ways, namely, periodic safety message (hereby called beacon) and event-driven message (hereby called emergency message), both sharing only one control channel. The beacon messages are messages containing status information about the sender vehicle, such as position, speed, heading, and others. Beacons provide fresh information about the sender vehicle to the surrounding vehicles in the network, updating them of the status of the current network and predicting the movement of vehicles. Beacons are sent aggressively to neighboring vehicles at 10 messages each second. In turn, this causes an increase in channel collision that the control channel cannot tolerate, especially when dense traffic occurs in small geographic areas. Therefore, it is necessary to formulate strategies to control the channel load resulting from packet collision and efficiently utilize the channel limited resources, especially during high dense vehicular traffic situations (Figure 1).

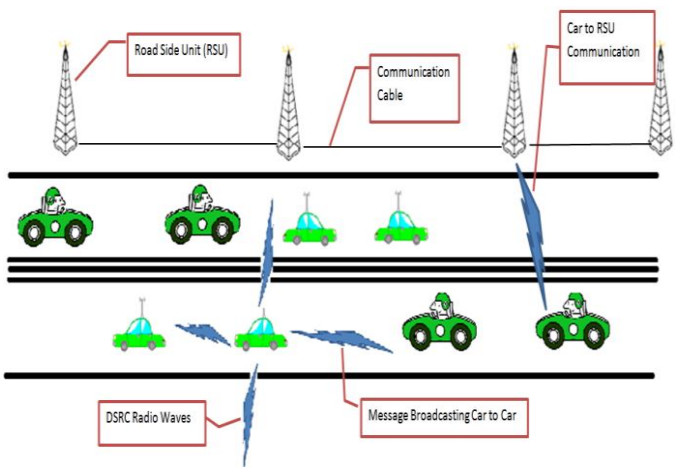

Figure 1: VANET Structure.

The VANET structure controlling beacon messages could be executed by transmission power control or message repetition control. Sending the message on high full power may cause the message to reach longer distances, thereby increasing the channel load, whereas sending in low power enables the message to reach only very short distances [9, 10].

To optimize and improve the channel performance, a dynamic transmission power control protocol is also proposed– called Particle swarm optimization Beacon Power Control (PBPC) – to adjust the transmission power of the beacon message that has been aggressively sent by all vehicles on the road at a frequency of 10 times/second.

## II. RELATED STUDIES

Power control in ad hoc networks has been an active topic for many years in the field of topology control. However, vehicular networks' main design goal as a safety system makes all these analyses or proposed algorithms insufficient in satisfying VANET requirements. Most of these studies addressed uni-cast environments and have been intended to improve energy consumption. In the literature, some studies have proposed the best path to the destination that minimize energy consumption and/or maximizes the overall throughput, including those of [11, 12, and 13].

### 1. Popular protocols:

In [14] authors have proposed an "energy aware" adaptive algorithm, which uses only local information to adjust power. [15, 16, and 17] all agree that the minimum transmission power does not always maximize throughput. Although many studies in this field can be found, VANET energy efficiency is not an issue where nodes have a nearly unlimited power supply for communication. In [18] authors proposed dynamic adjustment of transmission power based on estimates of local vehicle density. However, traffic density does not indicate channel load; thus, if the channel load is high and the traffic density is low, the sender chooses high power for sending the message, further increasing channel load and causing message reception failure [19].

In [20] authors presented a comparison between single-hop transmission at high transmission power and multi-hop transmission at low transmission power to determine whether or not efficient multi-hop beaconing can reduce channel load. The author found that single hop is best for beaconing and multi hop is best for full coverage. Sending in high power enables beacons to reach long distances in single-hop and may increase channel load. Broadcasting at full power, by comparison, produces a broadcast storm problem [21] and raises channel load.

Meanwhile, [22] developed a power control algorithm to determine optimum transmission power for beacon message transmission by adding a power tuning feedback beacon during each beacon message exchange. On each message exchange, the sender calculates the distance to the receiver and sets a predicted transmission power. On the receiver side, the distance is computed to determine if the transmission power achieved a greater distance or not. However, the delay resulting from these message exchanges makes the information gathered outdated as network status is variable.

In [23] authors proposed an analytical model to find a transmission power, which maximizes single-hop broadcast coverage. Authors in [24] also proposed an adaptive algorithm that adjusts to a given fixed transmission power. Although both studies focused on a pure broadcast environment, their assumptions made their approach infeasible for vehicular networks, because their nodes were static and had the same priority, i.e., there was no difference between the transmission power of beacon and emergency messages.

In [25] authors proposed a Delay-Bounded Dynamic Interactive Power Control (DB-DIPC), in which the transmission powers of VANET nodes are verified by neighboring vehicles at run-time. The idea is to send beacons to neighbor vehicles at very low power, and if the sender receives an acknowledgment, then that specific power is sufficient for close neighbors. This mechanism sends beacons to very close vehicles and limits the information gain for vehicles in the network. It also produces a very long delay as the sender needs to send the message many times to its neighbors and wait for a reply to decide the suitable transmission power.

### 2. The DFPAV Protocol

In [26] authors proposed the Fair Power Adjustment for Vehicular environments (DFPAV), which tries to adjust the channel load in a VANET environment by maximizing the minimum transmission range for all nodes using a synchronized approach. This is done by analyzing the piggybacked beacon information received from neighbors.

The adjusted power values are derived from Equation (1).

$$PA = \frac{MaxBeaconingLoad}{2 * CS_{max} * Vehicle\_density * Load_{vehicle}} - \epsilon \quad (1)$$

where PA is the power adjustment value, Max Beaconing Load is a pre-defined value and the beacon load must be kept below it, CS is the communication range for the vehicle, Vehicle density is the number of vehicles over 1,000 m, and Load is the current load of the vehicle messages.

The FPAV is part from the project Network On Wheels (NOW) [27] which is a German research project founded by DaimlerChrysler AG, BMW AG, Volkswagen AG, Fraunhofer Institute for Open Communication Systems, NEC Deutschland GmbH and Siemens AG in 2004. The project adopts an IEEE 802.11 standard for wireless access. The main objectives of this project are to solve technical issues related to communication protocols and data security for car-to-car communications. In this paper, the outcome of this project (the FPAV part) is adopted and compared with the proposed protocol of this paper.

The FPAV protocol is widely recognized for controlling channel load in a fair manner. In this scheme, every node uses a localized algorithm based on a "water filling" approach as proposed by [28] and starts transmitting the beacon message with the minimum transmission power. All the nodes increase their transmit power simultaneously to the same maximum power, while the constraint on the beaconing network load MBL is not violated.

Each node collects the received information, compares the maximum power reached by its neighbors, and then sends at a power value higher than the maximum reached by its neighbors. However, this fairness is not appropriate in a highly mobile network like VANET [29], where vehicles are always moving. This is because a road with smooth traffic can become heavily congested in a few seconds, and vehicles in heavy traffic areas cannot send at the same power as those in light areas. Moreover, each vehicle's signal depends on the surrounding vehicles and channel status, not on transmission power decided by vehicles 1,000 m away.

According to the analysis of the DFPAV protocol conducted by [20], the overhead for the existing DFPAV approach can be reduced, but there is still room for improvement.

Figure 2 shows the DFPAV flowchart.

In this paper, the DFPAV protocol is compared with the proposed PBPC protocol.

[30] based transmission range on traffic density estimation, in which an algorithm sets vehicle transmission range dynamically according to local traffic conditions. This protocol analyzes traffic conditions and not the channel status; hence, the channel may sometimes suffer from collisions when traffic is not dense.

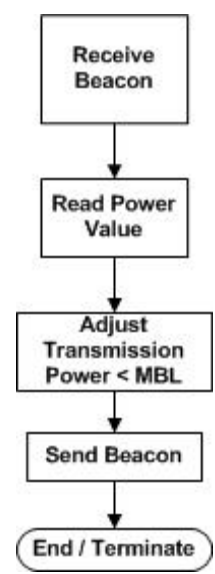

Figure 2: The DFPAV flowchart.

In [31] authors proposed power control assignment based on network channel busy time as wireless channel quality. When the channel busy time is higher or lower than a desired threshold, specific actions are conducted. However, since threshold selection is arbitrary, outcomes are not always optimal.

### III. PROPOSED PROTOCOL

This section presents a detailed design description for the PBPC protocol, which aims to control the beacon transmission power dynamically to decrease the packet collision resulting from the aggressive beacon sending in the channel. Decreasing the packet collision decreases the channel load, thereby enhancing the channel in such a way that it allows the beacons and the emergency messages to perform better.

### 1. Receiving Beacons

As mentioned earlier, each vehicle sends 10 beacons each second to its neighboring vehicles and suppose to receive the same number from each neighbor each second.

After receiving beacons from neighbors, a vehicle starts to collect the information gained and inserts it into neighbor table (NT) see table 1. NT contains the ID, position, speed, and direction of each vehicle sending the beacon. The information in the table is ordered so that neighbors close to the vehicle are prioritized; this also helps the vehicle draw the topology of the network around it. To avoid the increase in the channel collision, the channel status is tested before making any beacon transmissions. In [32] authors have proposed the method of computing channel status based on the beacons received in one second by the vehicle (Figure 3). From the figure, this vehicle received eight beacons from Vehicle A and six beacons from Vehicle B and so on. To compute the channel collision, Equation (2) is used.

TABLE 1: NEIGHBOR TABLE (NT).

| ID | Position(Latitude, Longitude point) | Speed (Km/H) | Direction |
|---|---|---|---|
| Vehicle A | 0.094449823, 1.751159661 | 120 | East |
| Vehicle B | 0.094449823, 1.761159661 | 80 | East |
| Vehicle C | 0.094449823, 1.759159661 | 65 | East |

| Vehicle A | 15 | 16 | 17 | 18 | | 20 | 21 | | 23 | 24 |
|---|---|---|---|---|---|---|---|---|---|---|
| Vehicle B | 71 | 72 | | | 75 | | | 78 | 79 | 80 |
| Vehicle C | 89 | 90 | | | | | | 96 | 97 | |
| Vehicle D | 22 | 23 | 24 | 25 | 26 | 27 | | 29 | 30 | |
| Vehicle E | 61 | 62 | 63 | | | | 67 | | 69 | 70 |

Figure 3: Sequence List (SL) Beacons received by a vehicle in 1 second.

$$CP = (1 - \frac{\sum B}{N \times 10}) \times 100\% , \qquad (2)$$

Where CP is the collision probability, B is the beacon received, and N is the number of neighboring vehicles.

For example the collision probability for the figure (3) is

$$CP = \left(1 - \frac{32}{5 \times 10}\right) \times 100\% = (1 - 0.64) \times 100\% = 36\%.$$

### 2. Channel analysis

Each beacon received provides information about neighboring vehicles and the current network. In the proposed protocol, beacons can decrease packet collision in the channel, thus increasing channel performance. Current beacon structure allows the addition of transmission power information to help the receiver determine the suitable power for transmission (Figure 2).

The power information added is piggybacked onto the current beacon used in the VANET. Each receiver vehicle keeps the sequence of received beacons in a Sequence List (SL) to determine the status of network traffic (Figure 5).

Given that the aim of this protocol is to reduce channel collision, the first step would be to measure the current packet collision. It should be noted that beacons only arrive if the network is not congested; if something prevents it from reaching its destination, the beacon fails. The percentage of collision can be computed by knowing the number of un-received beacons per second, as each vehicle must receive 10 beacons from each neighbor every second.

| Seq | Int | TS | Pos | Spd | Dir | PopBest | PowU |
|---|---|---|---|---|---|---|---|

Piggybacked information

Figure 4. Proposed beacon structure.

Where Seq: Beacon Sequence Number, Int: Beacon Interval, TS: Time Stamp (µ second), Dir: Direction Pos: Position, Spd: Speed (k/h), PopBest: Personal Best Power Achieved by the Beacon's Sender (dBm), PowU: Power Used By Sender (dBm).

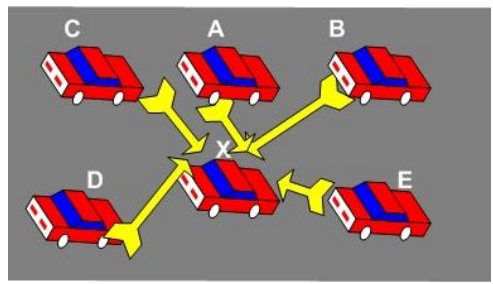

Figure 5: Vehicle X receives beacons from neighbors.

Vehicle X in the previous example analyzes the received beacons: from Vehicle A, the percentage of reception is 80%, meaning the percentage of failed beacons is 20%, with beacons 19 and 22 missing. Vehicle X also has to consider the distance between the two vehicles, as the percentage of received beacons drops as distance increases. Distance is obtained from the difference between the vehicle's current position and the position of the sender, D is the distance between vehicles.

Equation 3 computes the reception failure percentage for a single vehicle depending on the information provided by the SL, where p is the percentage of received beacons for a single vehicle.

$$f = \frac{(100-p)}{D} \qquad (3)$$

Returning to the last example, Vehicle A receives eight beacons in one second, and thus from Equation (2), cp=80%, and from (3), f = 1.538, indicating that 1.538 beacons fail for every meter. Therefore, if the distance is increased by one meter, 1.538 beacons would be lost and the percentage of received beacons would be 78.46%. The results are shown in Table 2, which shows the Distance Table (DT).

Table 2: Distance Table for Vehicle X.

| ID | Parentage of Reception | Distance | Fail |
|---|---|---|---|
| Vehicle A | 80 | 13 m | 1.538 |
| Vehicle B | 60 | 18 m | 2.22 |
| Vehicle C | 40 | 23 m | 2.6 |
| Vehicle D | 80 | 18 m | 1.11 |
| Vehicle E | 60 | 15 m | 2.667 |

Table 3: Active Beacon List (ABL).

| Seq | Int | TS (μs) | ELP | Position(Latitude point, Longitude point) | Speed (k/h) | Dir | PopBest (dBm) | PowU (dBm) |
|-----|-----|---------|-----|-------------------------------------------|-------------|-----|---------------|------------|
| 15 | 50 | 1.02 | A | 5.924449823, 7.51959661 | 60 | E | 28 | 25 |
| 71 | 50 | 1.02 | B | 5.924449823, 6.51759661 | 80 | E | 29 | 28 |
| 89 | 50 | 1.02 | C | 0.094049823, 1.755159661 | 70 | E | 28 | 29 |
| 22 | 50 | 1.02 | D | 0.094479823, 1.755159661 | 50 | E | 27 | 28 |
| 61 | 50 | 1.02 | E | 0.094479823, 1.755159661 | 70 | E | 26 | 28 |

The DT organizes the information about neighbors, including vehicle IDs, percentage of reception, and distance between sender and receiver.

In the beacon illustration in Figure 4, the receiver vehicle initiates a new list called the Active Beacon List (ABL), which includes all the information gained from neighboring vehicles, especially the power information (Table 3).

From the ABL, Vehicle X can analyze the transmission power for received beacons from its neighbors at any given moment. The received power depends on the distance between the two parties and on channel status. For instance, if Vehicle C transmits at power below 29 dBm, the transmitted beacon may not reach its destination, but if it expands the beacon's range by sending it at higher power, this may cause many more collisions.

Equation (4) takes the difference between the maximum and minimum power values received by the vehicle. These values are extracted from the ABL. A specific power value allows the beacon to reach its destination, but it may not be enough for the beacon to reach all adjacent neighbors. Since the maximum beacon power received contributes to the previously-computed collisions, the maximum power received must be decreased in order to reduce channel collisions. Minimum power, by contrast, must be increased to ensure that this beacon reaches farther neighbors, but not so much as to exceed the maximum power received. The decrease and increase in power values depends on the analysis obtained from Equation (9).

$$PD = MaxP - MinP. \quad (4)$$

Equation 2 computes the percentage of overall faults caused by all neighbors mentioned in DT, which in the last example is the 2.027% fault for each meter.

$$F = \sum_{n=1}^{n} \left( \frac{100-P}{D} \right) \div n \quad (5)$$

Meanwhile, Equation (6) computes the success percentage for beacon reception for the current neighborhood, which represents the parameter lBest as a fitness function for the PSO optimization algorithm.

$$S = (100 - \left( \frac{\sum_{n=1}^{n} D}{n} \times F \right))\% \quad (6)$$

Where n is the number of vehicles.

The computed success percentage indicates the effect of the power on the current channel, as not all power differences cause collisions in the channel. Equation (7) is used to compute the power that should be taken from the power difference, and the result becomes the lBest.

$$lBest = MinP + (PD \times lBestv). \quad (7)$$

The ABL provides the other parameters required for the PSO algorithm: pBest and gBest. The PSO algorithm is as follows (equation 8):

$$Sv = lBestv * w + C1 * rand1 * (pBestv - lBestv) + C2 * rand2 * (gBestv - lBestv). \quad (8), [33].$$

Equation (9) computes the optimal power for beacon transmission, which depends on channel analysis and successful transmission power used from neighboring vehicles. Afterwards, the vehicle inserts this information into the ready beacon, adjusts the transmission power to the newly computed value (PowU) and sends the beacon.

$$PowU = pBestv + Sv. \quad (9), [33].$$

Where W is random number between W: 0.1 to 0.5, C1= 2, C2= 2, rand: random number 0.1 to 1, pBest is the last lBest computed by the vehicle. w is the inertia weight of the particles, random 1 and random 2 are two uniformly distributed random numbers in the range [0, 1], and C1 and C2 are specific parameters which control the relative effect of the individual and global best particles.

lBest represents the success percentage resulting from the channel status analysis made by the current vehicle that, in turn, depends on the current channel status. Sometimes, the current vehicle makes a wrong analysis based on current channel readings, thus leading to wrong decisions. This shows that depending on just a single analysis may not lead to the best decision. In PSO, decision making depends on the best status that the vehicle has reached (lBest) and on the best analysis obtained by other vehicles (gBest). gBest can be computed from the ABL, where the highest result means the best analysis reached in the channel. pBest sent from neighboring vehicles represents its lBest, which is its analysis for the channel. Receiving more than one analysis helps the vehicle make better decisions (Figure 6), demonstrating how neighbors' performance influences the vehicle in making decisions using PSO optimization.

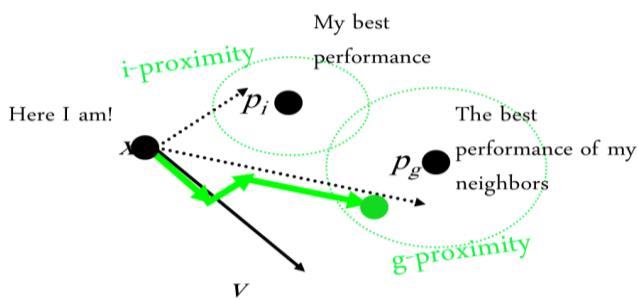

Figure 6: PSO optimization decision.

The "power used" sent through the beacon helps a vehicle determine the transmission power that enables the message to arrive at its current destination. It is also the optimal power resulting from the sender's analysis of channel status, giving the current vehicle a more comprehensive idea of a neighboring vehicle's analysis of the channel status. The PSO algorithm takes three input parameters (i.e., lBest, pBest, gBest) and produces the optimized lBest depending on channel analysis made by all neighboring vehicles and its own history. This provides a more accurate analysis and improved decision-making.

The resulting lBest must be processed to determine optimal transmission power. The following example shows how the previous equation works:

Equation (3) is performed to compute the beacon reception fault for each vehicle presented in the sequence list (Figure 3). The fault for Vehicle A, for example, can be computed as follows:

$$f = \frac{100-80}{13} = 1.538$$

The results of this equation are inserted in Table 2. Equation (4) computes the power difference between the maximum and minimum power received, where the maximum power contributes to the current channel collision and the minimum power is the least value power at which the receiver could receive messages. It is important to note that, below this value, the receiver may not receive the message. Deploying Equation (4) is as follows:

$$PD = 29 - 25 = 4$$

Afterwards, the vehicle computes the overall system fault for the channel (not for a specific vehicle) by performing Equation (4), in which total channel fault plays an important role in determining current packet collision.

$$F = \frac{10.135}{5} = 2.027$$

This channel fault is the opposite of channel success, and to compute for channel success, Equation (6) is performed:

$$lBest = S = 100 - (\frac{\sum D}{n} \, F))\%$$

$$lBest = 100 - (\frac{87}{5} \times 2.027) = 0.65$$

This means that the channel experiences collisions 35% of the time and is free of collision 65% of the time. The previously-computed power difference contributes to collision. Although not all power differences cause collisions, 35% of the power results in one. Therefore, to compute the best power value for the current channel status, Equation (7) is applied as follows:

$$lBest = MinP + (PD \times lBest) \qquad (7)$$

So the result will be:

$$lBest = 25 + (0.65 \times 4) = 27.6$$

This is the current power depending on the sender's channel analysis (lBest). Applying the PSO results in two parameters: the gBest and pBest. gBest is obtained from the Table 3, which shows that each vehicle in the network inserts its channel analysis (lBest) into the transmitted beacon, and the best analysis achieved by a vehicle's neighbors is taken and becomes gBest. pBest is obtained by the channel analysis history. Applying the PSO via Equation (8) can be done as follows:

$$27.6 \times 0.1 + 2 \times 0.7 \, (26 - 27.6) + 2 \times 0.6 \, (29 - 27.6)$$

$$= 2.76 - 2.24 + 1.68 = 2.20$$

$$PowU = lBest = 26 + 2.20 = 28.20, \text{ eq (9)}.$$

Therefore, the optimal transmission power for the beacon, based on channel analysis and PSO, is 28 dBm.

IV. SIMULATION

*Simulation Setup*

In order to test correctness of our protocol we made the simulation using the commercial program Matlab®, the distribution used is Nakagami distribution [34].

Parameters used in our simulation are summarized in table 4; all the simulations in this paper will adopt these parameters.

The simulation is made for 10s including 200 vehicles in 2km road consisting of 3 lanes.

Simulation Parameters

Parameters used in the simulation experiment are summarized in table 4; all the simulations in this paper will adopt these parameters.

V. RESULTS

In order to decrease the collision resulting from aggressive beacon broadcasting, a PBPC protocol is proposed and tested. The respective results of the experiments are listed below in Figures 7, 8, and 9. The implemented PBPC is compared with the performance of the NOW project's outcome, which is the DFPAV protocol.

TABLE 4: SIMULATION CONFIGURATION PARAMETERS

| Parameter | Value | Description |
|-----------|-------|-------------|
| Radio propagation model | Nakagami-m, m = 3 | Model m=3 is fixed as recommended in [26] |
| IEEE 802.11p data rate | 6Mbps | Fixed value |
| PLCP header length | 8 µs | Fixed value |
| Symbol duration | 8 µs | Fixed value |
| Noise floor | -99dBm | Fixed value |
| SNR | 10 - 40 dB | Adjustable to add noise to the signal |
| CW Min | 15 µs | Fixed value |
| CW Max | 1023 µs | Fixed value |
| Slot time | 16 µs | Fixed value |
| SIFS time | 32 µs | Fixed value |
| DIFS time | 64 µs | Fixed value |
| Message size | 512 bytes | Fixed value |
| Beacon Message Rate | 10 Message / s | Fixed value |
| Number of Vehicles | 200 | Fixed value |
| Road Length | 2 KM | Fixed value |
| Car Speed | 20km – 120km | Fixed value |
| Simulation Time | 10 s | Fixed value |
| Road Type | Highway | Fixed value |
| Number of lanes | 3 lanes | Fixed value |
| Neighbor entry size | 15 Bytes | Fixed value |

As mentioned earlier, the DFPAV depends on the fairness concept, where network vehicles have to share the same transmission power value and start to increase this power, while in PBPC each vehicle has to test and analyze the channel for collision and receive the best analyses achieved by neighbors. Afterwards, the sender vehicle applies the PSO to conclude the best transmission power for the current channel status. Figures 7 and 7 show how the protocol works and how the vehicle and the channel respond to beacon transmission power and channel collision changes, respectively. Figure 7 represents the average power used by the vehicle for beacon transmission. Every second, each vehicle analyzes the channel collision status presented in Figure 8 and adjusts the transmission power depending on the results of the analysis. Figure 8 shows the channel collision resulting from the beacon transmission. It is easy to notice that the impact of the power adjustment presented in Figure 7 is demonstrated by the channel collision presented in Figure 8.

As discussed earlier in this chapter, power value selection depends on a few factors, namely, the vehicle channel analysis, on its analysis history for best result reached, and on the neighboring vehicles analysis of the channel collision status. Meanwhile, for the DFPAV protocol, the transmission power starts at 25 dBm, and by the fair sharing method, all the vehicles in the network begin to raise the transmission power as long as the beaconing is less than a pre-defined Maximum Beaconing Load (MBL) value. This technique does not depend on the channel status but on a pre-defined threshold. For instance, in second two in Figure 8, the collision increases in such a way that the vehicle concludes that the power should be decreased to that used in the first second, as this value does not cause a high collision for the channel. This result depends on three parameters (i.e., lBest, gBest, pBest) as an input for the PSO optimizing technique. It is also worth noting that the selected power values do not make high changes as all the analysis are related and depend on the channel readings. High power changes affect channel collision and decrease channel performance.

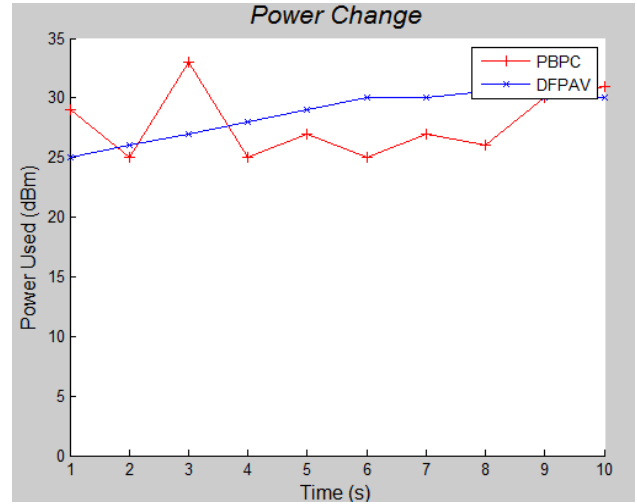

Figure 7: Average power used by the vehicles through the experiment.

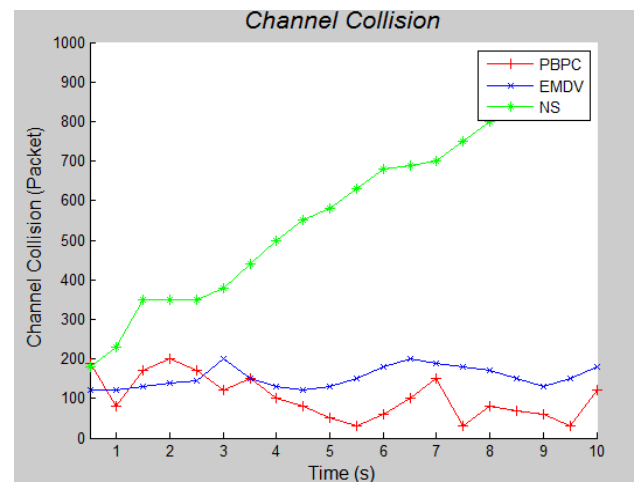

Figure 8: Collision resulted from power changing on beacons broadcasted

Table 5 shows the impact of a neighbor's analysis on the sender's decision. If the sender's analysis and the sender's last analysis results are high and the neighbor's analysis is low, then the sender slightly decreases the transmission power. This is because the sender cannot send in high power (even though its analysis calls for high power), because all of its neighbors tells it to send the beacon at low power.

TABLE 5: IMPACT OF NEIGHBOR VEHICLES ANALYSIS ON SENDER'S DECISION.

| lBest | pBest | gBest | Impact on Decision |
|-------|-------|-------|--------------------|
| High | High | High | Stay High |
| Low | High | High | Small Decrease on Power |
| Low | Low | High | Medium |
| Low | Low | Low | Low |
| High | Low | High | Increase Power |
| High | High | Low | Small Decrease on Power |

Figure 9 shows the effect of applying the PBPC protocol on the transmitted beacons. The result compares the delay resulting from the proposed protocol network with the DFPAV protocol and the NS, which is the VANET network working without deploying any protocol. The figure shows that the delay for PBPC is shorter than that for DFPAV by approximately 45%, as the channel scores better performance when deploying the PBPC, hence, shorter delay in the message broadcasting.

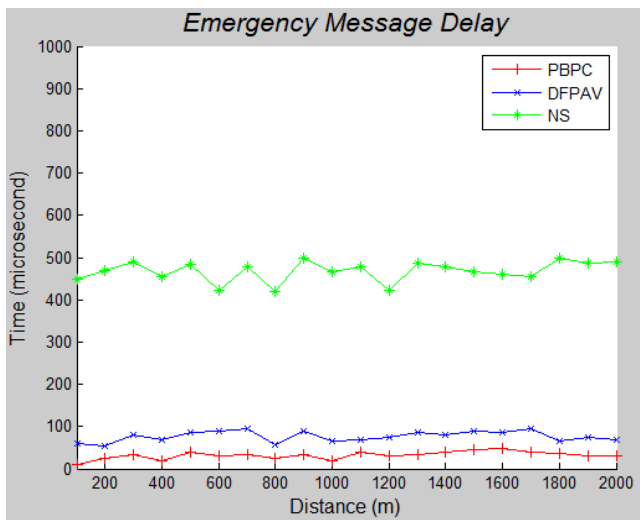

Figure 9: Beacon delay resulted from power adjustment.

VI. CONCLUSION

In this paper the PBPC (Particle swarm optimization Beacon Power Control) protocol is presented which limits the beaconing load on the channel depending on the channel status analysis concluded from the vehicles own analysis, and the best analysis resulted from neighboring vehicles making the error chance is very low while ensuring a high probability of beacon reception at close distances from the sender. Additionally, the performance of the protocol is evaluated in Matlab® environment.

ACKNOWLEDGMENT

This research is funded by the Deanship of Research and Graduate Studies in Zarqa University /Jordan.